\documentclass[conference]{IEEEtran}
\IEEEoverridecommandlockouts
\usepackage{cite}
\usepackage{amsmath,amssymb,amsfonts}
\usepackage{algorithmic}
\usepackage{graphicx}
\usepackage{textcomp}
\usepackage{xcolor}
\usepackage[misc]{ifsym}
\usepackage{multirow}
\usepackage{svg}
\usepackage{amsmath}
\usepackage{graphicx} 
\usepackage{epstopdf}
\def\BibTeX{{\rm B\kern-.05em{\sc i\kern-.025em b}\kern-.08em
    T\kern-.1667em\lower.7ex\hbox{E}\kern-.125emX}}
\begin{document}

\title{DDM: A Demand-based Dynamic Mitigation for SMT Transient Channels
\\
}

\author{\IEEEauthorblockN{Yue Zhang\textsuperscript{1,2}, Ziyuan Zhu\textsuperscript{1,2(\Letter)}, Dan Meng\textsuperscript{1}}
\IEEEauthorblockA{\textit{\textsuperscript{1}Institute of Information Engineering, Chinese Academy of Sciences} \\
\textit{\textsuperscript{2}School of Cyber Security, University of Chinese Academy of Science}\\
Beijing, China \\
\{zhangyue1995, zhuziyuan, mengdan\}@iie.ac.cn}
}

\maketitle

\begin{abstract}
Different from the traditional software vulnerability, the microarchitecture side channel has three characteristics: extensive influence, potent threat, and tough defense. The main reason for the micro-architecture side channel is resource sharing. There are many reasons for resource sharing, one of which is SMT (Simultaneous Multi-Threading) technology. In this paper, we define the SMT Transient Channel, which uses the transient state of shared resources between threads to steal information. To mitigate it, we designed a security demand-based dynamic mitigation (DDM) to Mitigate the SMT transient channels. The DDM writes the processes’ security requirements to the CPU register sets, and the operating system calls the HLT instruction to dynamically turn on and off the hyper-threading according to the register values to avoid the side channels caused by execution resource sharing. During the implementation of the scheme, we modified the Linux kernel and used the MSR register groups of Intel processor. The evaluation results show that DDM can effectively protect against the transient side-channel attacks such as PortsMash that rely on SMT, and the performance loss of DDM is less than 8\%.

\end{abstract}


\section{Introduction}
In recent years, computer security researchers have discovered many side channel attacks based on computer microarchitecture. These attack programs exploit the flaws of CPU microarchitecture design to steal secret information. The microarchitecture side channel is different from the traditional attack because those security risks come from underlying hardware. Besides, computers threatened by the vulnerabilities are widely distributed, nearly a decade of Intel, AMD and ARM CPU products, including most personal computers and cloud servers, are all under threat. How to solve micro-architecture side channels is a critical issue to protect computer system security.

The root cause of microarchitecture side channels is resource sharing. Many reasons lead to resource sharing, and SMT (Simultaneous Multi-Threading) is one of them. SMT, also called hyper-threading, is a computer parallel technique, which permits several independent threads to issue to multiple functional units each cycle. The objective of SMT is to substantially increase processor utilization in the face of both long memory latencies and limited available parallelism per thread [2]. In order to keep the implementation overhead of SMT below 5\% to the relative chip size and meet the maximum power requirements. Only the architecture state and some registers are duplicated, but that logical processors share nearly all other resources on the physical processor core, such as caches, execution units, branch predictors, control logic, and buses [21]. The shared resources mentioned above between threads (logical cores) in the same physical bring many side channels.

SMT transient channel is a kind of microarchitectural side channel that exploits the transient state of shared resources between threads in a shared physical core to steal key information. Transient state means the microarchitecture state utilized by side channels are not preserved during context switches. We will describe the specific details related to the above definition in Section II. PortsMash, CacheBleed, and MemJam are all examples of SMT transient channels.

In order to protect against the SMT transient channel, computer security researchers and engineers have come up with many defensive measures. Yarom Y [9] recommended to increase the cache bank bandwidth or modify the way of memory accesses to protect victim processes from CacheBleed threat. Percival [23] suggested partitioning the L1 cache between threads to eliminate the cache contention. However, new caches or new memory access only improve the security of future computers, not existing processors. In addition, these measures only address some transient channels that are caused by cache sharing within a core. In addition to cache sharing, SMT also presents security risks caused by execution port sharing [6] and ALU sharing [21]. Constant times [8,13,20] could theoretically solve all micro-architecture time side channels. Moghimi [22] proposed that only code with true constant-time properties, i.e., constant execution flow and constant memory accesses can be expected to have no remaining leakage on modern microarchitectures. However, constant time brings a significant performance overhead reduced its feasibility, sometimes source codes of application software also need to be modified. Disable SMT is one of the most common defense methods in the industry today, and it requires no changes to the processor hardware and software code. However, turning off hyperthreading can also cause significant throughput and performance losses. In order to keep SMT on, STEALTHMEM [24] did not assign the logical cores of the same physical core to different VM to avoid SMT channels. Although STEALTHMEM guarantees SMT on, it does not make full use of hyper-threading technology to improve server throughput, and it is only applicable to the server-side. The above defense methods are static. The static characteristic includes two aspects. Firstly, defense time is static so that the computer system has been under the influence of defense all the time. Secondly, defense space is static. All processor cores in computer system will be affected. 

The drawback of static defense is that it compromises the performance of the whole system in order to guarantee the security of a small number of processes. In order to reduce performance losses, Z Wang proposed a selective partitioning solution [25]. Selective partitioning suggested that the change of process scheduling or resources reallocation only works for critical processes, but they did not give the specific implementation scheme. Similar to selective partitioning, a system-level detection mechanism is proposed in [26], which monitors processes’ behaviors and matches its features with malicious programs. The operating system makes adjustments if the process behaves suspiciously. Requiring prior knowledge of malicious processes is a disadvantage of mechanism in [26]. 

To sum up, two problems exist in the current defenses: incomplete coverage of types and significant performance loss of static defense. Therefore, the purpose of this study is to improve the previous work and find a mechanism to mitigate the SMT transient channel. The defense time and space should be dynamic and based on users’ demand to decrease performance loss. In this work, we propose a low performance loss mechanism called DDM.

We evaluate the implementation and show that DDM effectively mitigates PortsMash, and we theoretically show that it can also defend other SMT transient channels. The performance evaluation shows that DDM introduces less than   8\% negligible performance overhead.

This paper makes the following contributions:
\begin{itemize}
\item This survey summarizes and proposes the concept of SMT transient channel (section II).
\item We use HLT instructions in a new way to adjust the processor (supporting hyperthreading) resources allocation.
\item We propose a new defense mechanism called DDM (section III) to defend against SMT transient channels and minimize performance losses. This mechanism schedules logical cores or threads dynamically based on the demands of the user’s applications.
\end{itemize}

We implement our new mechanism in section IV and make a security and performance evaluation of the new mitigation in section V. Discussion is covered in Section VI, and Section VII concludes.

\section{Background}

\subsection{Microarchitecture}
In order to improve the utilization of processor hardware resources, chip designers have been working on mining the potential parallelism of instruction execution. Superscalars do so by issuing and executing multiple instructions from a single thread, exploiting instruction-level parallelism [12]. In order to heighten instruction-level parallelism, multi-threaded superscalars hide latencies of one thread by switching to and executing instructions from another thread, thereby exploiting thread-level parallelism [3]. However, CPU will be limited by the instruction-level parallelism in a single thread. In SMT processors, thread-level parallelism can come from either multi-threaded, parallel programs or individual, independent programs in a multiprogramming workload [3]. SMT makes a single physical processor appear as multiple logical processors. Each logical processor has one copy of the architecture state, and the logical processors share a single set of physical execution resources [4].

Table 1 shows the internal resources of the SMT processor and their shared state. For details, please refer to [4].
\renewcommand{\arraystretch}{1.3}
\setlength{\tabcolsep}{8pt}
\begin{table}[htbp]
\caption{Sharing status of resources pipeline}
\begin{center}
\begin{tabular}{|c|c|c|}
\hline
\textbf{Pipeline}&{\textbf{Hard resources}}&{\textbf{Shared / Duplicated}} \\
\cline{1-3} 
{\multirow{2}*{IF}}& {ITLB}& {Duplicated} \\
\cline{2-3}
{}&{Branch prediction}&{Either duplicated or shared}\\
\cline{1-3}
{\multirow{2}*{ID}} & {TC}& {\multirow{2}*{Shared}} \\
\cline{2-2}
{}&{Microcode ROM}&{}\\
\cline{1-3}
{} & {Instruction Scheduling}& {} \\
\cline{2-2}
{EXE}&{Execution Units}&{Shared}\\
\cline{2-2}
{}&{Execution Port}&{}\\
\cline{1-3}
{} & {DTLB}& {} \\
\cline{2-2}
{MEM}&{Cache}&{Shared}\\
\cline{2-2}
{}&{Bus}&{}\\
\cline{1-3}
{Retire}&{-}&{Shared}\\
\cline{1-3}
\hline

\end{tabular}
\label{tab1}
\end{center}
\end{table}

\subsection{SMT Transient Channel}
The side-channel attack is a kind of physical attacks in which an adversary tries to exploit physical information leakages such as timing information, power consumption, or electromagnetic radiation [5] to get some secret message from victims. 

The micro-architecture time side channel is a kind of side channel. It uses the time-variant during program execution to infer critical information. Notably, the time track is related to micro-architecture state. Micro-architecture state includes persistent state and transient state [21]. So channels includes persistent state channel and transient state channel. Persistent state can be maintained during context switching, and persistent state channels exploit the limited storage space within the targeted microarchitectural resources. Transient state cannot be retained during the context switching. Transient-state channels, in contrast, exploit the limited bandwidth of the targeted element. 

SMT channel is a subset of the microarchitectural time side channels. Because of SMT technology, the attacker and the victim share microarchitectural components in a physical core. Attacker gets secret message through analysing the state of shared micro-architecture components. As a kind of microarchitectural time side channel, the SMT side channel is also divided into persistent and transient channels. The state of the microarchitecture utilized by the SMT transient state channel cannot be maintained in context switching, and the channel needs the support of Hyper-Threading Technology. Figure 1 describes the relationship between different side channels.


\begin{figure}
\centering
\includegraphics[width=0.5\textwidth]{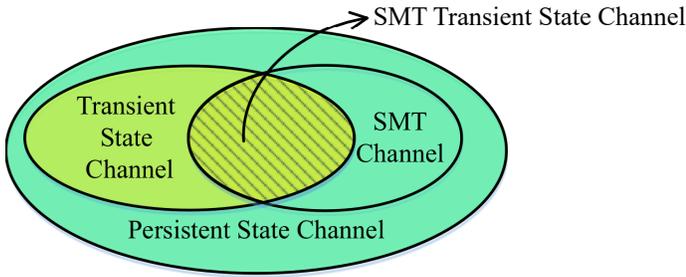}
\caption{The relationship between different side channels and the scope we research.}
\label{1}
\end{figure}

Typical SMT transient channels are PortsMash, Cachebleed, MemJam and so on. PortsMash was proposed by Aldaya et al in 2018 [6]. This side-channel attack exploits CPU execution port contention. If two threads are using the same port during execution, there will be CPU execution port contention. Although Instruction scheduling ensures fairness by alternating execution of instructions, resource contention leads to time delays. By recording and analyzing the time delay, attackers can infer key information indirectly. CacheBleed was presented by Yarom et al recently. It is a new cache-timing attack affecting some older processors featuring Hyper-Threading such as Sandy Bridge. The authors exploit the fact that cache banks can only serve one request at a time, thus issuing several requests to the same cache bank, i.e. accessing the same offset within a cache line, results in bank contention, leading to timing variations and leaking information about low address bits [9]. MemJam is a side-channel attack that exploits false dependency of memory read-after-write and provides a high-quality cache level timing channel.

\subsection{Threat Model}
Figure 1 shows that this scheme only considers the protection scheme for SMT transient channel, and we do not consider other side-channel attacks, including cross-core attacks. Shared resources related to SMT transient channel include cache (L1 Cache, L2 Cache), BTB, execution port, function unit, ALU. It contains all the shared resources associated with SMT, not just one of them.

\section{Demand-based Dynamic Mitigation}
To mitigate the SMT transient channel, we propose the DDM mechanism. As shown in Figure 3, the entire mechanism consists of two paths: demand path and defense path.

\subsection{Demand path}
The demand path writes the user's requirements to the logical core registers, including security and performance demands. As shown in Figure 3, the start node of the demand path is the input of the user's demand, which goes through the interface application and operating system successively, and finally writes the user’s demands into the maps in corresponding logical cores. Here we describe the role and internal principles of steps in the demand path in turn.

\paragraph{Start Node (User demands)}Users' demand is the requirement for the execution environment of their process. Here, we consider the users' security demands and performance demands. Security demands (SD): The security requirements of users on the execution environment reflects the importance of user information. The more important the information contained in the user process, the more secure the execution environment is required. Performance demand (PD): Performance demand describes the upper limit of user tolerance for processor performance losses. Specifically, it is a limitation on the amount of performance loss that can be caused by securing critical processes.

\paragraph{Second node (Interface Software)}Interface software is a tool, which provides an interface for users to write their requirements into the computer system. Users run the interface software and enter their security and performance demand as prompted.

\paragraph{Third node (OS)}After receiving the call of interface software, the operating system calls the system function to write/read the requirements of the users into/from the hardware register of each logical core.

\paragraph{End node (MAPs)}DDM establishes two maps inside each logical core. The two maps are demand map and action map respectively. As shown in Figure 2, demand map establishes a mapping relationship between users’ identity and their requirements. After writing their own identity and demands into maps, the system can query user requirements according to the user ID. Action map establishes the mapping relationship between user requirements and defense measures.We assign an Action ID(AD) to each defense. The system can query the corresponding defense measures according to requirements.

\begin{figure}[h]
\centering
\includegraphics[width=0.4\textwidth]{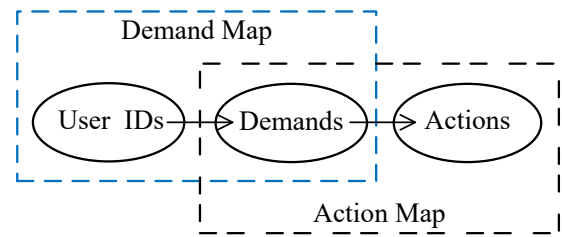}
\caption{End node (MAPs)}
\end{figure}

\begin{figure*}
\centering
\includegraphics[width=0.75\textwidth]{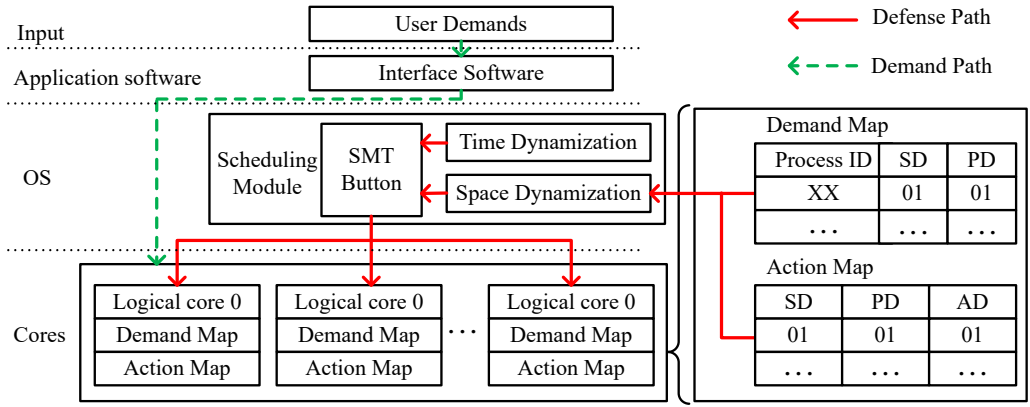}
\caption{Structure of DDM}
\end{figure*}

\subsection{Defense Path}
The task of defense path is to make corresponding dynamic defense measures according to user requirements. As shown in Figure 3, the start node of the defense path is security and performance maps in every logical core. The next step is to control the operation of SMT button with the output information of the space dynamization component and the time dynamization comp onent (both are sub-component of Scheduling module). SMT button aims to defend the SMT transient state channel by minimizing resource sharing among logical cores in the same core through process scheduling. We introduce the role and internal principle of each node in the defense path in turn.

\paragraph{Start Node (MAPs)}Read the content in maps to get the user's security and performance demands.

\paragraph{Second Node (Space Dynamization)}Dynamic defense space determines the scope of defensive actions based on data in maps, such as whether to defend the entire computer system or protect only a single logical core. Because SMT-based side-channel attacks mostly occur between different logical cores of the same physical core (we call them homologous logical core (HLC)), processes running on the HLC are the critical defense objects. The space dynamization component needs to determine the HLC of the logical core on which the critical program resides. The basic granularity of the defense is the logical core, so we first need to bind the key processes to a fixed logical core.

\paragraph{Third Node (Time Dynamization)}The role of this node is to determine the running time of defense measures. To minimize the performance penalty of suspending execution resource sharing, the processor needs to minimize the hang time of the logical core as much as possible. The basic idea is to implement defense during the execution of key processes. The system reverts to its original state before and after critical process execution. Time dynamization component tracks the execution time of key processes.

\paragraph{Fourth Node (SMT Button)}
In order to minimize the shared resources between threads (logical cores) of the same physical core during critical process execution. SMT button sets the logical core to HALT state temporarily.

Currently, most of the existing methods disable SMT via BIOS setting. Some methods with the help of the operating system to use the system command to turn off hyper-threading of all physical cores. In order to reduce the performance loss, this paper proposes a new method of turning off SMT, which has two characteristics: first, it can selectively turn off the SMT function of the specified physical core; second, we can restart the thread before critical thread starting and after it finishing.

This new approach relies on the processor's low-power switching capabilities. There is an instruction in Intel's instruction set called HLT. The instruction can stop instruction execution and places the processor in a HALT state. An enabled interrupt, a debug exception or a signal will resume execution. If an interrupt is used to resume execution after an HLT instruction, the saved instruction pointer (CS:EIP) points to the instruction following the HLT instruction[10]. Especially, when an HLT instruction is executed on an Intel 64 or IA-32 processor supporting Intel Hyper-Threading Technology, only the logical processor that executes the instruction is halted. The other logical processors in the physical processor remain active unless they are each individually halted by executing an HLT instruction. 

\begin{figure}[h]
\centering
\includegraphics[width=0.5\textwidth]{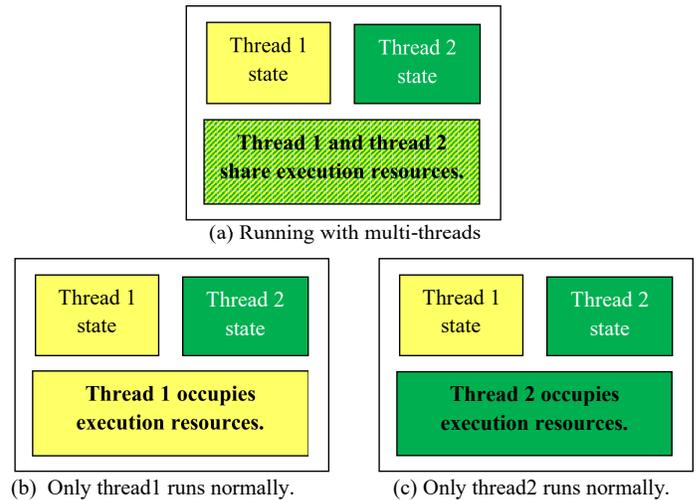}
\caption{On a processor with Hyper-Threading Technology, resources are allocated to a single logical processor if the other threads are in halt state. When all logical cores run at the same time, resources are shared between all threads.}
\end{figure}

\begin{figure*}[h]
\centering
\includegraphics[width=\textwidth]{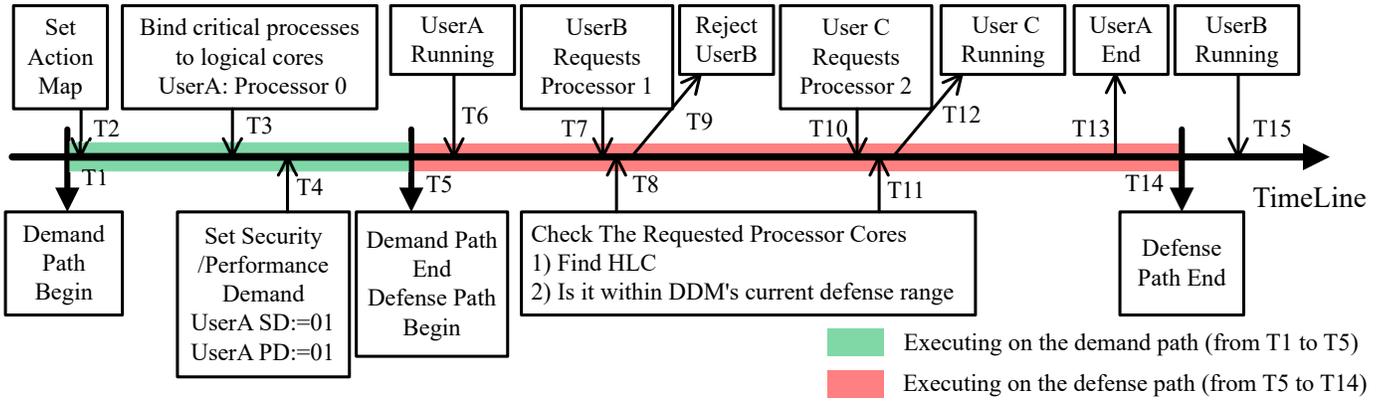}
\caption{Application of DDM in specific scenarios}
\end{figure*}

As shown in Figure 4(a), when multiple logical cores are running simultaneously on a physical core, the unit that stores processor state (mostly registers) is unique to each logical core, and the execution resources are shared. As shown in Figure 4(b) and 4(c), When only one thread (one logical processor) is active, all execution resources are re-combined to give the single active thread (logical processor). HLT stops the execution of the instruction but does not change the state of the register. Therefore, when the logic core is stopped to a low-power state, it does not affect the execution of other logic cores or interferes with the execution state of its own thread, and it also eliminates the side-channel attack caused by execution resource sharing.

In a real scenario, when a critical thread needs to be isolated, the system will suspend the other logical cores in some physical into a low-power state. When the critical program finished, the recovery signal is sent, and the rest instructions continue to execute. In essence, this new approach does not really disable SMT but rather suspends the logical core or thread execution dynamically. However, it achieves the same result as shutting down SMT directly, avoiding contention for execution resources.

\paragraph{End Node (Logical Cores)}The logical core is scheduled by SMT button. If there are two logical cores in a physical core, one of which executes the key program and applies for protection, then the other logical core will automatically switch to HALT state once receiving the scheduling signal of SMT button. Then the resource shared between threads will be distributed to only a logical core, so that SMT transient state channel will be avoided.

\subsection{Illustrative Example}
The example in Figure 5 demonstrates a usage scenario on a processor with DDM. The DDM execution timeline has two phases: Executing on the demand path (from T1 to T5) and Executing on the defense path (from T5 to T14).

\begin{itemize}
\item At T2, the processor administrator completes the setting of the processor action map and determines the relationship between user demands and security measures.
\item At T3, DDM binds the user process to the specified processor logic core. At T4, users write their security and performance requirements to the processor. Take user A as an example, assuming that user A is a protected critical process. User A is scheduled to execute on processor 0, his security requirement id is 01, and his performance requirement id is 01, both of which are written to the register of processor 0. According to the Action map written in advance, the security and performance requirements of user A jointly determine that DDM will take the defensive measures coded as 01 to protect user A.
\item At T6, user A starts running on the DDM-protected processor.
\item At T7, user B requests to run, and then at T8, DDM checks it. The results show that user B's operation may threaten the security of user A's process. Because user B will be running on processor 1, which belongs to the same physical core as processor 0, there will be some sharing of storage and execution resources. So at T9, processor 1 blocks user B from running until user A finishes.
\item Unlike user B, user C passed the security check. Because HLC of processor 2 did not execute the critical task. Therefore the execution of user C does not threaten the information security of critical programs, and user C is up and running at T12.
\item At T13, user A task finishes and DDM turns off the defense switch. The computer system revert to the raw operating state.
\item After the DDM is turned off, user B executes normally on processor 1 on T15.
\end{itemize}

\section{Implementation}
Table II summarizes the experimental environment for our experiment and evaluation. We use the Intel (R) Core(TM) i7-6700 processor from the Skylake architecture family with a 3.4GHZ main frequency. The CPU supports SMT and has four physical cores. Each physical core is divided into two logical cores and supports the execution of two threads at the same time. The operating system used in the experiment was Ubuntu 16.04. Similar to the vast majority of processors supporting hyper-threading, most resources in Skylake architecture processor are fully shared between threads to improve the dynamic utilization of the resources, including L2 and LLC caches and all the execution units[10]. The experimental results are representative.

\renewcommand{\arraystretch}{1.3}
\setlength{\tabcolsep}{10pt}
\begin{table}[htbp]
\caption{Experimental environment}
\begin{center}
\begin{tabular}{|c|c|}
\hline
{Processor model}&{Intel (R) Core(TM) i7-6700}\\
\cline{1-2}
{Microarchitecture}&{Skylake}\\
\cline{1-2}
{Clock frequency}&{3.4GHZ}\\
\cline{1-2}
{Physical cores}&{4}\\
\cline{1-2}
{Logical cores}&{8}\\
\cline{1-2}
{Kernel version}&{Kernel-4.17.4 (Ubuntu16.04)}\\
\cline{1-2}
{SMT support}&{support}\\
\cline{1-2}
\hline
\end{tabular}
\end{center}
\end{table}

\subsection{Implementation of Demand path}
\paragraph{Core Level}DDM writes the demand map and action map to register groups. In the implementation of the scheme, we use the MSR register group of Intel CAT. MSR (Model Specific Register) is a concept in x86 architecture. It refers to a series of registers used in x86 architecture processors to control CPU operation, function switch, debugging, trace program execution, monitor CPU performance and other aspects. When it came to the Intel Pentium processor, Intel officially introduced RDMSR and WRMSR, two instructions for reading and writing MSR register. Figure 6 shows the implementation details of the two maps in the MSR registers. 

\begin{figure}[h]
\centering
\includegraphics[width=0.4\textwidth]{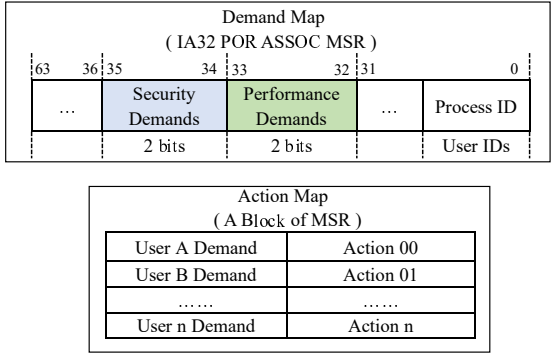}
\caption{Implementation of MAPs}
\end{figure}

We quantified the security demands (SD) of users and used 2 bits to describe the SD of users. The larger the number, the higher the security requirement. Two bits are also used to quantify the user's performance demands (PD). The larger the code, the greater the performance loss. As shown in Figure 6, the security demands correspond to the 34 to 35 bit of the MSR register. The performance demands correspond to the 32 to 33 bit of the MSR register. The lower bits of the register store the user process ID. The action map is established by a block of MSRs (the number of MSRs is security and performance demands combinations).

\paragraph{Application Software level}To read and write registers, we made a tool called DDMtool to finish the task of interface software. Users can use it to read and write MSR register sets directly.

\subsection{Implementation of Defense path}
Most implementations of the defense path are in the operating system.
\paragraph{Space Dynamization}
We determine the space of defenses based on the user's security and performance demands. Different requirements correspond to different defense intensity, and different defense intensity corresponds to different defense space. The weakest way coded Action 00 is to do nothing to protect the processor. The second coded Action 01 is to protect only a single physical core. The logical cores suspended by SMT buttons belong to only one physical core. The strongest coded Action 10 is to protect all processor cores. In other words, suspend all logical cores except the one used by the user who needs protecting.
We define a variable CD whose value is equal to the geometric average of SD and PD:

$${CD=\sqrt{SD^{2}+PD^{2}}}$$
When CD satisfies $CD<\sqrt{2}$,we take action00. When CD satisfies ${\sqrt{2}}\leq{CD}\geq2{\sqrt{2}}$, we take action01. The same way, if CD satisfies $CD>2{\sqrt{2}}$, we take action10.

\paragraph{Time Dynamization}The processor monitors the operation of the key process and its logical core. When the key process runs, the SMT button opens; when the key process finishes, the SMT button closes.
\paragraph{SMT Button}We utilize HLT instruction to added a system call function called sys\_hlt( ) with system call number 548. The logical core that calls this system call will halt,avoiding contention between logical cores for execution resources.

\section{Evaluation}
\subsection{Security Evaluation}
To evaluate the security, We tested to see if DDM can defend CPU against PortsMash attacks.We targeted a tunnel TLS server authenticating with a P-384 certificate, the Spy measuring port contention in parallel as the server produces ECDSA signatures. Due to the application of hyper-threading technology, there is contention for execution ports. PortsMash captures the port contention delay during double and add operations, resulting in an accurate raw signal trace containing the sequence of operations during scalar multiplication, and leaking enough information of secret nonce k to later succeed in the key recovery phase.

\begin{figure*}[htp]
\centering
\includegraphics[width=0.9\textwidth]{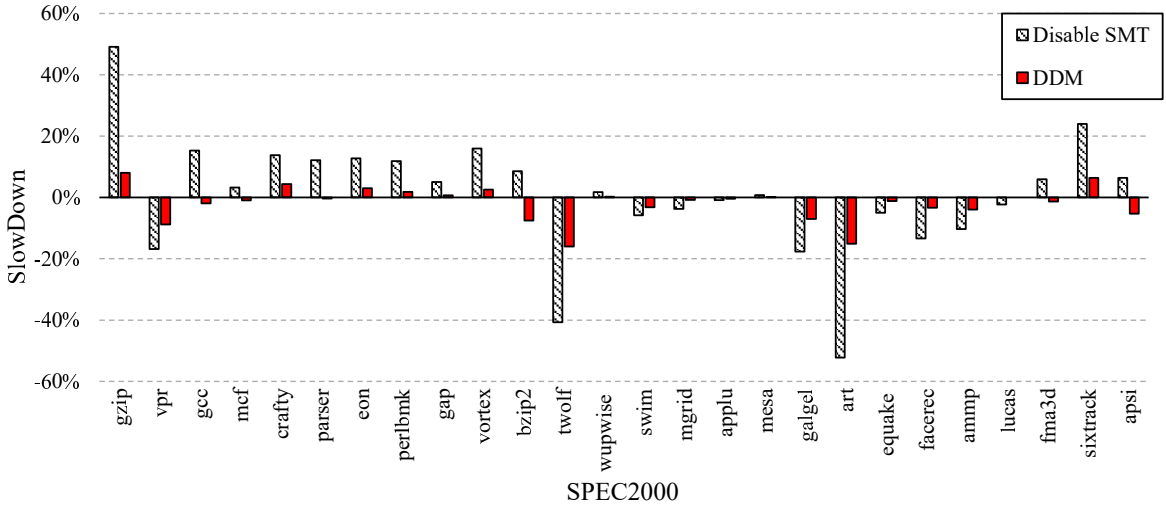}
\caption{The average slowdown of each of SPEC2000 benchmarks with two different mitigation. }
\end{figure*}

\begin{figure}[h]
\centering
\includegraphics[width=0.5\textwidth]{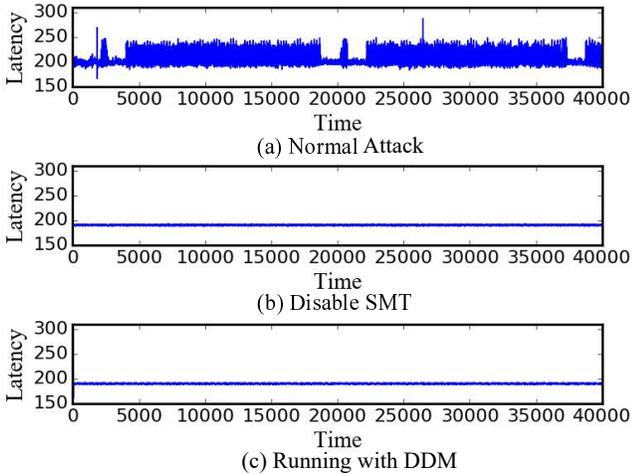}
\caption{The time delay tracks observed by spies in different situations. (a) is a normal attack time delay diagram without adding any defensive measures. (b) is the time delay trace detected by spy after SMT is turned off in BIOS Settings. (c) is the time delay track obtained by the key correlation program running under the protection of DDM.}
\end{figure}

Figure 8 shows the time delay tracks observed by spies in different situations. Figure 8(a) is no protection for ECDHE and ECDSA, and an attacker can easily detect the regular delay trajectory caused by the execution port contention. Then the mathematical operations related to the key are deduced from the time delay trajectory. However, when using the DDM defense measures shown in Figure 8(c), it is difficult for an attacker to detect the delay changes caused by resource contention. As Figure 8(b) shows, we know that turning off SMT in BIOS can successfully protect against PortsMash, then the time delay trace is smooth. When the system is under DDM protection, the program execution time delay trace is also smooth. Both time traces in figure 8(b) and figure 8(c) are independent of keys. Therefore, we think that DDM can effectively protect against PortsMash and it is impossible to recover private key by spying ports contention.

\subsection{Performance Evaluation}
We measure the impact on legacy applications using SPEC2000. We run each benchmark three times to completion using the reference (for SPEC2000) input sets and compare the results against the baseline cases without the DDM and without hyper-threading. Because the DDM takes up CPU running time, we expect some performance degradation. We choose SPEC rate metrics to measure the throughput or rate of CPU carrying out a number of tasks[18]. The experimental processor has 8 logical cores, so we set 8 copies of the benchmarks that are running simultaneously during the experiment. Figure 7 shows that the performance of the DDM slowdown is less than the closed hyper-threading directly. Evaluation results showed that the performance loss caused by DDM was less than 8\% in both int and fp tests. It is worth noting that some values in the test were negative numbers when hyperthreading was turned off, indicating improved performance. DDM does not cause performance loss in these benchmarks but brings performance gain to the processor. Our explanation for this experimental phenomenon is that DDM dynamically shuts down hyperthreading, reducing the congestion caused by resource competition and shortening the running time of benchmarks such as 179.art and 300.twolf. To sum up, we conclude that DDM incurs a small performance degradation.

\section{Discussion}
This study shows that DDM is a practical, simple and low performance loss method to mitigate SMT-based side channels. It has been demonstrated in two ways. First, in our security assessment, we have demonstrated that DDM effectively protects against PortsMash vulnerabilities caused by execution port contention. In addition, because DDM can turn off execution resource sharing dynamically between logical cores during the execution of a process, it theoretically protects the processor from most side-channels that rely on hyper-threading execution resource sharing. Second, in the performance evaluation, experimental results show that DDM has a small performance loss. Especially measuring and comparing compute-intensive integer(CINT) performance, the throughput or rate of a processor with DDM has a big advantage over processor which is disabled SMT statically in bios.

We have attempted to extend two aspects of previous work on contralateral channel defense. The first is to expand the scope of defense against side-channel attacks, rather than limiting them to cache side channels. The second is to propose a dynamic low-performance loss strategy, which is different from the previous high-performance loss defense schemes such as constant time[13, 14, 15, 19, 20], disable SMT[11] and inject noise[16].

\section{Conclusion}
We propose DDM, a lightweight system security mechanism for the cloud provider and cloud customers to protect security-sensitive code and data against execution resources shared side channels, by tuning the hyper-threading of the processor dynamically. DDM builds on existing commodity hardware and can be easily deployed. It bridges the gap between protecting security applications and the existing performance-oriented hyper-threading technology.

The DDM uses the MSR register set to establish two mappings within the processor, which are security demands and security actions. Users can write security requirements for code and data into the map. Information written to registers is managed and invoked by the operating system. According to the safety requirements of users, we suspend logic cores dynamically to achieve the purpose of temporarily shutting down hyper-threading technology dynamically. Restore the normal function of the processor immediately after the critical program execution. Our evaluation shows that DDM not only effectively mitigates the SMT-based side channels but also introduces very small performance degradation.

\section*{Acknowledgment}
This work was supported by the Institute of Information Engineering (IIE), Chinese Academy of Sciences. Thanks to my family and friends for their support and understanding.


\vspace{12pt}

\end{document}